\documentclass[11pt]{article}

\usepackage{pgfplots}
\usepackage{tikz}
\usepackage{tikzpeople}
\usepackage{circuitikz}
\usetikzlibrary{tikzmark, shapes, fit,backgrounds}
\usetikzlibrary{matrix, decorations.pathreplacing, angles, quotes}
\usetikzlibrary{patterns, calc, positioning}
\usetikzlibrary{arrows, arrows.meta}
\usetikzlibrary{shapes.multipart}
\usetikzlibrary{fit}
\usetikzlibrary{shapes.geometric, positioning}
\usetikzlibrary{decorations.pathmorphing}
\tikzstyle{block} = [rectangle, draw, 
    minimum width=4em, text centered, rounded corners, minimum height=1em]
\tikzstyle{line} = [draw, -latex]
\tikzset{meter/.append style={draw, inner sep=10, rectangle, font=\vphantom{A}, minimum width=30, scale=.7, path picture={\draw[black] ([shift={(.1,.3)}]path picture bounding box.south west) to[bend left=50] ([shift={(-.1,.3)}]path picture bounding box.south east);\draw[black,-{Latex[scale=.5]}] ([shift={(0,.1)}]path picture bounding box.south) -- ([shift={(.3,-.1)}]path picture bounding box.north);}}}
\tikzset{snake it/.style={decorate, decoration=snake}}

\usepackage[utf8]{inputenc}
\usepackage{amsmath}
\usepackage{amssymb}
\usepackage{amsthm}
\usepackage{color,soul}
\usepackage{tikz}
\usetikzlibrary{topaths,calc}
\usepackage{subfigure}
\usepackage{cite}

\usepackage{braket}

\def\QED{\mbox{\rule[0pt]{1.5ex}{1.5ex}}}
\def\proof{\noindent\hspace{2em}{\it Proof: }}

\def\QED{\mbox{\rule[0pt]{1.5ex}{1.5ex}}}

\newtheorem{theorem}{Theorem}

\newtheorem{lemma}{Lemma}

\newtheorem{remark}{Remark}

\newcommand\blfootnote[1]{%
  \begingroup
  \renewcommand\thefootnote{}\footnote{#1}%
  \addtocounter{footnote}{-1}%
  \endgroup
}

\title{The Entropy Characterization of Quantum MDS Codes}
\author{Hua Sun}
\date{}

\begin{document}
\maketitle

\blfootnote{
Hua Sun (email: hua.sun@unt.edu) is with the Department of Electrical Engineering at the University of North Texas.}

\begin{abstract}
An $[[n,k,d]]$ quantum maximum-distance-separable code maps $k$ source qudits to $n$ coded qudits such that any $n-(d-1)$ coded qudits may recover all source qudits and $n = k + 2 (d-1)$. The entropy of the joint state of the reference system of $k$ qudits and the $n$ coded qudits is fully characterized - the joint state must be pure, i.e., has entropy zero; and any sub-system whose number of qudits is at most half of $k+n$, the total number of qudits in the joint state must be maximally mixed, i.e., has entropy equal to its size.
\end{abstract}

\newpage
\allowdisplaybreaks
\section{Introduction}

An $[[n,k,d]]_q$ quantum error correcting code encodes a quantum message $Q_0$ of $k$ source qudits into $n$ coded qudits $Q_1, \cdots, Q_n$, where each source/coded qudit is $q$-dimensional, such that from any $n-(d-1)$ coded qudits, the $k$ source qudits can be perfectly recovered. The well known quantum Singleton bound \cite{Knill_Laflamme, Rains, Cerf_Cleve, Grassl_Huber_Winter} states that 
\begin{eqnarray}
k \leq n - 2(d-1) \label{singleton}
\end{eqnarray}
i.e., for fixed number of coded qudits (code length) $n$ and fixed erasure correcting capability $d-1$ ($d$ is called minimum distance and can be defined equivalently through error correcting capability or codeword weights), the maximum number of source qudits $k$ allowed is upper bounded by $n - 2(d-1)$. A code that achieves the quantum Singleton bound with equality is called a quantum maximum-distance-separable (MDS) code 
and quantum MDS codes are well known to exist for any $k, d, n = k + 2(d-1)$ for sufficiently large prime power $q$ (see e.g., \cite{Grassl_Beth_Roetteler}). 
We focus exclusively on quantum MDS codes in this note.

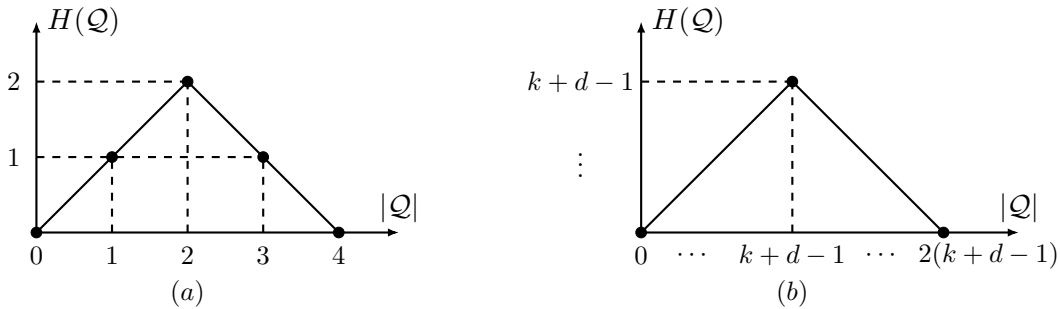
\begin{figure}[h]
\center
\begin{tikzpicture} 
\begin{scope}[shift={(0,0)}]
\draw [-{Latex[length=1.5mm]}, thick] (0,0)--(4.8,0) node [above] {$|\mathcal{Q}|$};
\draw [-{Latex[length=1.5mm]}, thick] (0,0)--(0,2.8) node [right] {$H(\mathcal{Q})$};
\draw [thick] (0,0)--(2,2)--(4,0);
\draw[thick, dashed] (0,2)--(2,2);
\draw[thick, dashed] (2,0)--(2,2);
\draw[thick, dashed] (1,0)--(1,1);
\draw[thick, dashed] (3,0)--(3,1);
\draw[thick, dashed] (0,1)--(3,1);
\node at (0,-0.3) {\small $0$};
\node at (1,-0.3) {\small $1$};
\node at (2,-0.3) {\small $2$};
\node at (3,-0.3) {\small $3$};
\node at (4,-0.3) {\small $4$};
\node at (-0.3,1) {\small $1$};
\node at (-0.3, 2) {\small $2$};
\filldraw[] (0,0) circle (2pt);
\filldraw[] (1,1) circle (2pt);
\filldraw[] (2,2) circle (2pt);
\filldraw[] (3,1) circle (2pt);
\filldraw[] (4,0) circle (2pt);

\node at (2, -0.8) {\small $(a)$};
\end{scope}

\begin{scope}[shift={(8,0)}]
\draw [-{Latex[length=1.5mm]}, thick] (0,0)--(5,0) node [above] {$|\mathcal{Q}|$};
\draw [-{Latex[length=1.5mm]}, thick] (0,0)--(0,2.8) node [right] {$H(\mathcal{Q})$};
\draw [thick] (0,0)--(2,2)--(4,0);
\draw[thick, dashed] (0,2)--(2,2);
\draw[thick, dashed] (2,0)--(2,2);
\node at (0,-0.3) {\small $0$};
\node at (0.7,-0.3) {\small $\cdots$};
\node at (2,-0.3) {\small $k+d-1$};
\node at (3.2,-0.3) {\small $\cdots$};
\node at (4.6,-0.3) {\small $2(k+d-1)$};
\node at (-0.8,1) {\small $\vdots$};
\node at (-0.8, 2) {\small $k+d-1$};
\filldraw[] (0,0) circle (2pt);
\filldraw[] (2,2) circle (2pt);
\filldraw[] (4,0) circle (2pt);
\node at (2, -0.8) {\small $(b)$};
\end{scope}
\end{tikzpicture} \label{fig}
\vspace{-0.1in}
\caption{The entropy value of any sub-system of $RQ_1\cdots Q_n$ for any $[[n, k, d]]$ quantum MDS code. $(a)$ $k=1, d=2$ and $(b)$ general $k,d$.} 
\end{figure}

\vspace{0.1in}
We aim to understand the Von Neumann entropy of any sub-system of the $n$ coded qudits $Q_1, \cdots, Q_n$, i.e., $H(Q_\mathcal{I})$ for any set $\mathcal{I} \subset \{1, 2, \cdots, n\} \triangleq [n]$ (where the entropy is measured in $q$-ary units, i.e., the base of logarithm is set as $q$). It turns out that the answer is particularly clean when we include the reference system 
$R$ in the picture and consider the joint state $RQ_1\cdots Q_n$ where $R$ is maximally entangled with 
$Q_0$ such that $R Q_0$ is a pure state and as a result, $R$ is maximally mixed and contains $k > 0$ $q$-dimensional qudits. Our objective now becomes to characterize $H(\mathcal{Q})_\rho$ where $\mathcal{Q} \subset \{R, Q_1, \cdots, Q_n\}$ and $\rho$ denotes the density matrix of the joint state $RQ_1\cdots Q_n$ (and is omitted in the subscript of entropy thereafter). For any $\mathcal{Q}$, we define $|\mathcal{Q}|$ as the number of qudits in $\mathcal{Q}$; for example, when $\mathcal{Q} = \{Q_1, Q_2\}$, $|\mathcal{Q}| = 2$ as each $Q_i$ has $1$ qudit and when $\mathcal{Q} = \{R, Q_1\}$, $|\mathcal{Q}| = k+1$ as $R$ has $k$ qudits. Interestingly, for any $[[n = k+2(d-1),k,d]]_q$ quantum MDS code, we show that $H(\mathcal{Q}) = \min(|\mathcal{Q}|, 2(k+d-1) - |\mathcal{Q}|)$ (see Fig.~\ref{fig}), i.e., the joint state must be pure, $H(R,Q_1,\cdots,Q_n) = 0$; and $H(\mathcal{Q}) = |\mathcal{Q}|$ when the size of $\mathcal{Q}$ is at most $k+d-1 = (k+n)/2$, half of the total size of the joint state. 
Therefore, the entropy characterization of any quantum MDS code is unique and the entropy value of any sub-system only depends on how many qudits the sub-system contains.

\vspace{0.1in}

\begin{theorem}\label{thm}
For any $[[n,k,d]]_q$ quantum MDS code ($n = k+2(d-1)$), the entropy of any sub-system of the reference system\footnote{In this work, we consider the most general coding model, where the source system $R Q_0$, along with some ancilla $Q_{anc}$ (of $a \geq n-k$ qudits) passes through a unitary transformation to the coded system $R Q_1 \cdots Q_n$, along with some auxiliary output $Q_{aux}$ (of $a - (n-k) \geq 0$ qudits) so that $R Q_1 \cdots Q_n Q_{aux}$ is a pure state ($R$ goes through an identity mapping). Note that some prior work (e.g., \cite{Cerf_Cleve} and Section II of \cite{Grassl_Huber_Winter}) considered a more specialized model where $Q_{anc}$ contains exactly $a = n-k$ qudits and $Q_{aux}$ does not exist. It turns out that interestingly, for quantum MDS codes, from Theorem \ref{thm}, $R Q_1 \cdots Q_n$ must be pure and $Q_{aux}$ must be in a product state with $R Q_1 \cdots Q_n$ so that there is no loss to not consider $Q_{aux}$, i.e., set $a = n-k$.} and coded qudits $RQ_1\cdots Q_n$ is given as
\begin{eqnarray}
H(\mathcal{Q}) = \min( |\mathcal{Q}|, 2(k+d-1) - |\mathcal{Q}| ). \label{entropy}
\end{eqnarray}
\end{theorem}

\subsection{Related Work}
We first write out (\ref{entropy}) more explicitly. For an index set $\mathcal{I} \subset [n]$, $|\mathcal{I}|$ denotes its cardinality and $Q_{\mathcal{I}}$ denotes the set of $Q_i$ such that $i \in \mathcal{I}$.
\begin{eqnarray}
&& H( Q_\mathcal{I} ) = |\mathcal{I}|, ~\mbox{when}~ |\mathcal{I}| \leq k + d - 1 \label{q1} \\
&& H( Q_\mathcal{I} ) = 2(k+d-1) - |\mathcal{I}|, ~\mbox{when}~ |\mathcal{I}| > k + d - 1 \label{q2} \\
&& H(R, Q_\mathcal{I}) = k + |\mathcal{I}|, ~\mbox{when}~ |\mathcal{I}| \leq d - 1 \label{r1} \\
&& H(R, Q_\mathcal{I}) = k + 2(d - 1) - |\mathcal{I}|, ~\mbox{when}~ |\mathcal{I}| > d - 1. \label{r2}
\end{eqnarray}
(\ref{q1}), (\ref{q2}) are previously known (from algebraic instead of entropic arguments and implicit in the proof of Theorem 8 in \cite{Huber_Grassl}).
\cite{Alsina_Razavi} notices that $R Q_1 Q_2 Q_3$ is an absolutely maximally entangled (AME) state (i.e., $H(\mathcal{Q}) = \min(|\mathcal{Q}|, 4 - |\mathcal{Q}|)$ where $\mathcal{Q} \subset \{R, Q_1, Q_2, Q_3\}$) for a $[[3,1,2]]_3$ quantum MDS code construction; this result is covered by Theorem \ref{thm}. More generally, AME states \cite{Facchi_Florio_Parisi_Pascazio, Helwig_Cui, Goyeneche_Alsina_Latorre_Riera_Zyczkowski} require each component of the joint state to have equal size (e.g., $1$ qudit), so Theorem \ref{thm} indicates that $R Q_1 \cdots Q_n$ is an AME state when $k = 1$. When $k > 1$, $R Q_1 \cdots Q_n$ can be viewed as a generalization of AME states where each component is not restricted to have equal size (the first component $R$ has $k>1$ qudits while all other components $Q_i$ have $1$ qudit each) but any bipartition is still maximally entangled. From this angle, Theorem \ref{thm} reveals a new entropic connection between quantum MDS codes and AME states, adding to various known connections from the literature in terms of code/state constructions, weight distributions etc. \cite{Huber_Grassl, Scott, Alsina_Razavi, Raissi_Gogolin_Riera_Acin}.

\section{Proof of Theorem \ref{thm}}\label{proof}
We prove the alternative form (\ref{q1}), (\ref{q2}), (\ref{r1}), (\ref{r2}) and it turns out that the order of consideration is crucial. First, we prove (\ref{q1}). From any $n - (d-1)$ coded qudits $Q_{\mathcal{I}}, \mathcal{I} \subset [n], |\mathcal{I}| = n - (d-1)$, we may recover the $k$ source qudits. The entropic condition for perfect recovery is proved by Schumacher and Nielsen \cite{Schumacher_Nielsen},
\begin{eqnarray}
2H(R) = I(R; Q_\mathcal{I}), ~\forall \mathcal{I} ~\mbox{where}~|\mathcal{I}| = n - (d-1) \label{dec}
\end{eqnarray}
whose intuitive meaning is that all entanglement between the reference system $R$ and the source message $Q_0$ (where entanglement is captured by mutual information between $Q_0$ and its purifying system $R$ being $2H(R)$) must be preserved in the entanglement between $R$ and any $n-(d-1)$ coded qudits. As a consequence, no information shall be leaked to the environment \cite{Schumacher_Nielsen}, i.e., the coded qudits that might be erased $Q_{\mathcal{I}^c}$ where $\mathcal{I}^c \triangleq [n] \setminus \mathcal{I}$ denotes the complement of $\mathcal{I}$ and for two sets $\mathcal{A}, \mathcal{B}$, $\mathcal{A} \setminus \mathcal{B}$ denotes the different set, i.e., the set of elements that belong to $\mathcal{A}$ but not to $\mathcal{B}$.
\begin{eqnarray}
I(R; Q_\mathcal{I}) + I(R; Q_{\mathcal{I}^c}) &=& 2H(R) - ( H(R \mid Q_\mathcal{I}) + H(R \mid Q_{\mathcal{I}^c}) ) \\
&\leq& 2H(R) \label{e1} \\
\overset{(\ref{dec})}{\Longrightarrow}~~~~~ I(R; Q_{\mathcal{I}^c}) &=& 0, ~\forall \mathcal{I} ~\mbox{where}~ |\mathcal{I}| = n - (d-1) \label{sec}
\end{eqnarray}
where (\ref{e1}) follows from weak monotonicity of quantum entropy functions \cite{Nielsen_Chuang, Pippenger}, i.e., $H(R \mid Q_\mathcal{I}) + H(R \mid Q_{\mathcal{I}^c}) \geq 0$. To obtain (\ref{sec}), note that $ I(R; Q_{\mathcal{I}^c}) \leq 0$ implies $ I(R; Q_{\mathcal{I}^c}) = 0$ as quantum mutual information is non-negative \cite{Nielsen_Chuang, Pippenger}. 

Consider any set $\mathcal{J}$ such that $\mathcal{J} \subset \mathcal{I}$ and $|\mathcal{J}| = d-1$. As $|\mathcal{J}^c| = n - (d-1)$, from (\ref{sec}) we have
\begin{eqnarray}
I(R; Q_\mathcal{J}) = 0 \label{sec1}
\end{eqnarray}
and next let us revisit the decoding constraint (\ref{dec}),
\begin{eqnarray}
2k &=& 2H(R) \label{} \label{rq} \\
&\overset{(\ref{dec})}{=}& I(R; Q_\mathcal{I})  =  I(R; Q_\mathcal{J}, Q_{\mathcal{I} \setminus \mathcal{J}})  \\
&=& I(R; Q_\mathcal{J}) + I(R; Q_{\mathcal{I} \setminus \mathcal{J}} \mid Q_\mathcal{J}) \\
&\overset{(\ref{sec1})}{=}& H ( Q_{\mathcal{I} \setminus \mathcal{J}} \mid Q_\mathcal{J}) - H(Q_{\mathcal{I} \setminus \mathcal{J}} \mid Q_\mathcal{J}, R) \\
&\leq& H ( Q_{\mathcal{I} \setminus \mathcal{J}} ) + H ( Q_{\mathcal{I} \setminus \mathcal{J}} ) \label{e2}\\
&\leq& 2 \sum_{i \in \mathcal{I} \setminus \mathcal{J}} H(Q_i) \label{ind2} \\
&\leq& 2 |\mathcal{I} \setminus \mathcal{J}| \label{dim} \\
&=& 2[ n - (d-1) - (d-1) ] \label{singleton} \\
&=& 2k \label{mds}
\end{eqnarray}
where (\ref{rq}) follows from the fact that the reference system $R$ is maximally mixed. In (\ref{e2}), the first term follows from the non-negativity of quantum mutual information, i.e.,
\begin{eqnarray}
0 &\leq& I ( Q_\mathcal{J} ; Q_{\mathcal{I} \setminus \mathcal{J}} ) \label{ind} \\
&=& H ( Q_{\mathcal{I} \setminus \mathcal{J}} ) - H ( Q_{\mathcal{I} \setminus \mathcal{J}} \mid Q_\mathcal{J}) 
\end{eqnarray}
and the second term follows from the Araki-Lieb inequality (also known as triangle inequality) for quantum entropy (which is a special case of weak monotonicity) \cite{Nielsen_Chuang, Pippenger}. (\ref{ind2}) also follows from the non-negativity of quantum mutual information, similar to (\ref{ind}). (\ref{dim}) is due to the dimension bound of quantum entropy, i.e., 
each $q$-dimensional qudit $Q_i$ may contain at most $1$ $q$-ary unit of quantum entropy,
\begin{eqnarray} 
H(Q_i) \leq \log_q q = 1, ~\forall i \in [n]. \label{qi}
\end{eqnarray}
(\ref{singleton}) gives us the quantum Singleton bound $k \leq n - 2(d-1)$ and for quantum MDS codes, the bound is tight, i.e., $n = k + 2(d-1)$ and (\ref{mds}) is obtained (see also the derivation in Chapter 7.8.3 of \cite{Preskill} and Chapter 12.4.3 of \cite{Nielsen_Chuang}). As the left-hand-side and right-hand-side of (\ref{mds}) are both $2k$, all the inequalities from (\ref{rq}) to (\ref{qi}) must be equalities. Specifically, (\ref{ind}) indicates that for any $\mathcal{K}_1, \mathcal{K}_2 \subset [n]$ such that $\mathcal{K}_1 \cap \mathcal{K}_2 = \emptyset, |\mathcal{K}_1 | \leq k, |\mathcal{K}_2| \leq d-1$, $I( Q_{\mathcal{K}_1} ; Q_{\mathcal{K}_2} ) = 0$. This is seen as follows. 
Consider any $\mathcal{J} \subset \mathcal{I}$ such that $|\mathcal{J}| = d-1, |\mathcal{I}| = n-(d-1) = k+d-1, \mathcal{K}_2 \subset \mathcal{J} \subset \mathcal{I}, \mathcal{K}_1 \subset (\mathcal{I} \setminus \mathcal{J})$.
\begin{eqnarray}
0  &\overset{(\ref{ind})}{=}& I ( Q_\mathcal{J} ; Q_{\mathcal{I} \setminus \mathcal{J}} ) = I ( Q_\mathcal{J} ; Q_{\mathcal{K}_1}, Q_{\mathcal{I} \setminus (\mathcal{J} \cup \mathcal{K}_1)} ) \\
&\geq& I ( Q_\mathcal{J} ; Q_{\mathcal{K}_1} ) \label{sub} = I ( Q_{\mathcal{K}_2}, Q_{\mathcal{J} \setminus \mathcal{K}_2} ; Q_{\mathcal{K}_1} ) \\
&\geq&I ( Q_{\mathcal{K}_2} ; Q_{\mathcal{K}_1} )  \label{sub1} \\
\Longrightarrow ~~~~ 0 &=& I ( Q_{\mathcal{K}_1} ; Q_{\mathcal{K}_2} )  = H(Q_{\mathcal{K}_2}) - H ( Q_{\mathcal{K}_2} \mid Q_{\mathcal{K}_1} )\label{ind1}
\end{eqnarray}
where (\ref{sub}) follows from the fact that conditional quantum mutual information is non-negative, i.e., $I ( Q_\mathcal{J} ; Q_{\mathcal{I} \setminus (\mathcal{J} \cup \mathcal{K}_1)} \mid Q_{\mathcal{K}_1}) \geq 0$ (equivalent to strong subadditivity/sub-modularity of quantum entropy \cite{Nielsen_Chuang, Pippenger}). (\ref{sub1}) is similar to (\ref{sub}). (\ref{ind1}) says that any at most $k$ coded qudits and any disjoint at most $d-1$ coded qudits are in a product state.
With a similar proof to (\ref{ind1}), (\ref{ind2}) being equality leads to that for any $\mathcal{K} \subset [n], |\mathcal{K}| \leq k$,
\begin{eqnarray}
H(Q_\mathcal{K}) = \sum_{i \in \mathcal{K}} H(Q_i) \label{ind3}
\end{eqnarray}
which says that any at most $k$ coded qudits are in a product state.

We are now ready to prove (\ref{q1}). For any $\mathcal{I} \subset [n], |\mathcal{I}| \leq n - (d-1) = k + d - 1$, suppose $\mathcal{I} = \{i_1, \cdots, i_{|\mathcal{I}|}\}$ and we use (\ref{ind1}) repeatedly to split $H(Q_\mathcal{I})$ into blocks of entropy of $k$ coded qudits,
\begin{eqnarray}
H(Q_\mathcal{I}) &=& H(Q_{i_1}, \cdots, Q_{i_k}) + H(Q_{i_{k+1}}, \cdots, Q_{i_{|\mathcal{I}|}} \mid Q_{i_1}, \cdots, Q_{i_k}) \\
&\overset{(\ref{ind1})}{=}& H(Q_{i_1}, \cdots, Q_{i_k})  + H(Q_{i_{k+1}}, \cdots, Q_{i_{|\mathcal{I}|}} ) \label{e3}\\
&\overset{(\ref{ind1})}{=}& H(Q_{i_1}, \cdots, Q_{i_k})  + H(Q_{i_{k+1}}, \cdots, Q_{i_{2k}} ) + \cdots \\
&\overset{(\ref{ind3})}{=}& H(Q_{i_1}) + \cdots + H(Q_{i_k}) + H(Q_{i_{k+1}}) + \cdots + H(Q_{i_{2k}}) + \cdots + H(Q_{i_{|\mathcal{I}|}}) \label{e4} \\
&\overset{(\ref{qi})}{=}& |\mathcal{I}|
\end{eqnarray} 
where (\ref{e3}) follows from setting $\mathcal{K}_1 = \{i_1, \cdots, i_k\}, \mathcal{K}_2 = \{i_{k+1}, \cdots, i_{|\mathcal{I}|}\}$  in (\ref{ind1}) and (\ref{e4}) follows from (\ref{ind3}) through setting $\mathcal{K}$ as $\{i_1, \cdots, i_k\}$, $\{i_{k+1}, \cdots, i_{2k} \}$ etc. The proof of (\ref{q1}) is thus complete.
\vspace{0.1in}

Second, we prove (\ref{r1}) as an immediate consequence of (\ref{q1}) and (\ref{sec}). For any $\mathcal{I} \subset [n], |\mathcal{I}| \leq d-1$, (\ref{sec}) indicates that $R$ and $Q_\mathcal{I}$ are in a product state.
\begin{eqnarray}
H(R, Q_\mathcal{I}) &\overset{(\ref{sec})}{=}& H(R) + H(Q_\mathcal{I}) \\
&\overset{(\ref{q1})}{=}& k + |\mathcal{I}|
\end{eqnarray}
and the proof of (\ref{r1}) is complete.
\vspace{0.1in}

Third, we prove (\ref{r2}). Consider any $\mathcal{I} \subset [n]$, $|\mathcal{I}| > d-1$ and depending on whether $|\mathcal{I}|$ is no greater than $k+d-1$ or not, we divide into two cases below. 
\begin{enumerate}
\item $|\mathcal{I}| \leq k+d-1$

Consider first $|\mathcal{I}| = k+d-1 = n - (d-1)$. From the decoding constraint (\ref{dec}), we have
\begin{eqnarray}
H(R, Q_\mathcal{I}) &=& H(R) + H(Q_\mathcal{I}) - I(R; Q_{\mathcal{I}}) \\
&\overset{(\ref{dec})}{=}& H(Q_\mathcal{I}) - H(R) \\
&\overset{(\ref{q1})}{=}& |\mathcal{I}| - k = d-1 = k + 2(d-1) - |\mathcal{I}| \label{r2e}
\end{eqnarray}
and (\ref{r2}) is proved for the case where $|\mathcal{I}| = k+d-1$. Next combining with the fact that from (\ref{r1}), for any $\mathcal{K} \subset \mathcal{I}$, $|\mathcal{K}| = d-1$,  $H(R, Q_\mathcal{K}) = k + |\mathcal{K}| = k+d-1$, we proceed to the immediate case of $H(R, Q_\mathcal{J})$ where $\mathcal{K} \subset \mathcal{J} \subset \mathcal{I}$ and $d-1 < |\mathcal{J}| \leq k+d-1$. For any such $\mathcal{J}$, we have on the one hand (connecting to $H(R, Q_\mathcal{K})$)
\begin{eqnarray}
H(R, Q_\mathcal{J}) &=& H(R, Q_\mathcal{K}, Q_{\mathcal{J} \setminus \mathcal{K}}) \\
&=& H(R, Q_\mathcal{K}) + H(Q_{\mathcal{J} \setminus \mathcal{K}} \mid R, Q_\mathcal{K}) \\
&\overset{(\ref{r1})}{\geq}& k+|\mathcal{K}| - H(Q_{\mathcal{J} \setminus \mathcal{K}}) \label{e5} \\
&\overset{(\ref{q1})}{=}& k+|\mathcal{K}| - |\mathcal{J} \setminus \mathcal{K}| \\
&=& k + |\mathcal{K}| - ( |\mathcal{J}|  - |\mathcal{K}|) \\
&=& k + 2(d-1) - |\mathcal{J}| \label{r2u}
\end{eqnarray}
where (\ref{e5}) follows from the Araki-Lieb inequality, i.e., $H(Q_{\mathcal{J} \setminus \mathcal{K}} \mid R, Q_\mathcal{K})  \geq - H(Q_{\mathcal{J} \setminus \mathcal{K}})$; and on the other hand (connecting to $H(R, Q_\mathcal{I})$)
\begin{eqnarray}
H(R, Q_\mathcal{J}) &=& H(R, Q_\mathcal{I}) - H(Q_{\mathcal{I} \setminus \mathcal{J}} \mid R, Q_{\mathcal{J}} ) \\
&\leq& H(R, Q_\mathcal{I}) + H(Q_{\mathcal{I} \setminus \mathcal{J}}) \label{e6}\\
&\overset{(\ref{r2e}) (\ref{q1})}{=}& d-1 + |\mathcal{I} \setminus \mathcal{J}| \\
&=& d-1 + |\mathcal{I}| - |\mathcal{J}| \\
&=& k + 2(d-1) - |\mathcal{J}| \label{r2l}
\end{eqnarray}
where (\ref{e6}) follows from the Araki-Lieb inequality. Combining the matching upper bound (\ref{r2u}) and lower bound (\ref{r2l}), we have proved
\begin{eqnarray}
H(R, Q_\mathcal{J}) = k + 2(d-1) - |\mathcal{J}|, ~\forall \mathcal{J} ~\mbox{where}~d-1 < |\mathcal{J}| \leq k+d-1 \label{r21}
\end{eqnarray}
and replacing $\mathcal{J}$ by $\mathcal{I}$ gives us the desired (\ref{r2}).

\item $|\mathcal{I}| > k+d-1$

We prove that (\ref{r2}) holds when $k+2(d-1) \geq |\mathcal{I}| \geq k+d-2$ through induction on $|\mathcal{I}|$ (while we are only interested in the case where $|\mathcal{I}| > k+d-1$ here, we include the two cases where $|\mathcal{I}| = k+d-1$ and $|\mathcal{I}| = k+d-2$ to use the established result (\ref{r21}) as the base case). The base case where $|\mathcal{I}| = k+d-2$ or $k+d-1$ has been proved in (\ref{r21}). 
Next we proceed to the induction step, i.e., we assume that (\ref{r2}) holds when $|\mathcal{I}| = k+d-2 + \Delta$ for any integer $\Delta$ such that $0\leq \Delta < d-1$ and prove that (\ref{r2}) holds when $|\mathcal{I}| = k+d-2 + \Delta+2$. Suppose $\mathcal{I} = \{i_1, i_2, \cdots, i_{k+d-2+\Delta+2}\}$. Then by the induction assumption, we have
\begin{eqnarray}
H(R, Q_{\mathcal{I} \setminus \{i_1\}}) &=& k + 2(d-1) - |\mathcal{I} \setminus \{i_1\}| = k + 2(d-1) - |\mathcal{I}| + 1 \label{e7} \\
H(R, Q_{\mathcal{I} \setminus \{i_2\}}) &=& k + 2(d-1) - |\mathcal{I} \setminus \{i_2\}| = k + 2(d-1) - |\mathcal{I}| + 1 \label{e8} \\
H(R, Q_{\mathcal{I} \setminus \{i_1, i_2\}}) &=&  k + 2(d-1) - |\mathcal{I} \setminus \{i_1, i_2\}| = k + 2(d-1) - |\mathcal{I}| + 2. \label{e9} 
\end{eqnarray}
On the one hand, by sub-modularity of quantum entropy functions we have 
\begin{eqnarray}
H(R, Q_{\mathcal{I} \setminus \{i_1\}})  + H(R, Q_{\mathcal{I} \setminus \{i_2\}}) &\geq& H(R, Q_{\mathcal{I} \setminus \{i_1, i_2\}}) + H(R, Q_{\mathcal{I}}) \\
\overset{(\ref{e7}), (\ref{e8}), (\ref{e9})}{\Longrightarrow} ~~~~~ H(R, Q_{\mathcal{I}}) &\leq& k + 2(d-1) - |\mathcal{I}| \label{r2uu}
\end{eqnarray}
and on the other hand, by the Araki-Lieb inequality we have
\begin{eqnarray}
H(R, Q_{\mathcal{I}}) &=& H(R, Q_{\mathcal{I} \setminus \{i_1\}}) + H(Q_{i_1} \mid R, Q_{\mathcal{I} \setminus \{i_1\}}) \\
&\overset{(\ref{e7})}{\geq}& k + 2(d-1) - |\mathcal{I}| + 1 - H(Q_{i_1}) \\
&\overset{(\ref{qi})}{=}& k + 2(d-1) - |\mathcal{I}| + 1 - 1 \label{e10} \\
&=& k + 2(d-1) - |\mathcal{I}| \label{r2ll}
\end{eqnarray} 
where to obtain (\ref{e10}), note that for quantum MDS codes, (\ref{qi}) must take equality. Now combining the matching upper bound (\ref{r2uu}) and lower bound (\ref{r2ll}), we have proved that (\ref{r2}) holds for any $\mathcal{I}$ where $|\mathcal{I}| > k+d-1$.

In particular, when $|\mathcal{I}| = k+2(d-1) = n$, (\ref{r2}) becomes 
\begin{eqnarray}
H(R, Q_1, \cdots, Q_n) = 0 \label{pure}
\end{eqnarray}
i.e., $RQ_1\cdots Q_n$ is a pure state.

\end{enumerate}
Combining the above two cases, we have completed the proof of (\ref{r2}).
\vspace{0.1in}

Fourth and finally, we prove (\ref{q2}) which is straightforward based on what has been established. We give two proofs here. The first proof uses the fact that $RQ_1\cdots Q_n$ is pure (refer to (\ref{pure})) and the property of a pure state (which follows from the Araki-Lieb inequality) that
\begin{eqnarray}
H( Q_{\mathcal{I}} ) &=& H(R, Q_{\mathcal{I}^c})  \\
(\mbox{when}~ |\mathcal{I}| > k+d-1) &\overset{(\ref{r1})}{=}& k + |\mathcal{I}^c| \\
&=& k + n - |\mathcal{I}| = 2(k+d-1) - |\mathcal{I}|
\end{eqnarray}
and the proof is complete. The second proof uses the decoding constraint (\ref{dec}). For any $\mathcal{I} \subset [n], |\mathcal{I}| > k+d-1$, $Q_\mathcal{I}$ can recover the source qudits\footnote{This can be proved entropically by picking any $\mathcal{J} \subset \mathcal{I}, |\mathcal{J}| = k+d-1$, then $2H(R) \geq I(R; Q_\mathcal{I}) \geq I(R; Q_\mathcal{J}) \overset{(\ref{dec})}{=} 2H(R)$ such that (\ref{dec1}) holds.} so that
\begin{eqnarray}
I(R; Q_\mathcal{I}) &=& 2H(R) \label{dec1} \\
\Longrightarrow ~~~~ H(Q_\mathcal{I}) &=& 2H(R) + H(R, Q_\mathcal{I}) - H(R) \\
&\overset{(\ref{r2})}{=}& H(R) + k + 2(d-1) - |\mathcal{I}|  \\
&=& 2(k+d-1) - |\mathcal{I}|
\end{eqnarray}
and the proof of (\ref{q2}) is complete.

\begin{remark}
An intuitive explanation of (\ref{q1}), (\ref{q2}), (\ref{r1}), (\ref{r2}) is given as follows. $H(\mathcal{Q})$ monotonically increases with $|\mathcal{Q}|$ when $|\mathcal{Q}| \leq k + d - 1$ as here the system needs to be in a product state so as to contain sufficient information about the source qudits; while when $|\mathcal{Q}| > k+d-1$, $H(\mathcal{Q})$ monotonically decreases with $|\mathcal{Q}|$ as here the system needs to be sufficiently entangled to ensure the source qudits can be recovered.
\end{remark}

\bibliography{Thesis}

\begin{thebibliography}{10}
\providecommand{\url}[1]{#1}
\csname url@samestyle\endcsname
\providecommand{\newblock}{\relax}
\providecommand{\bibinfo}[2]{#2}
\providecommand{\BIBentrySTDinterwordspacing}{\spaceskip=0pt\relax}
\providecommand{\BIBentryALTinterwordstretchfactor}{4}
\providecommand{\BIBentryALTinterwordspacing}{\spaceskip=\fontdimen2\font plus
\BIBentryALTinterwordstretchfactor\fontdimen3\font minus
  \fontdimen4\font\relax}
\providecommand{\BIBforeignlanguage}[2]{{%
\expandafter\ifx\csname l@#1\endcsname\relax
\typeout{** WARNING: IEEEtran.bst: No hyphenation pattern has been}%
\typeout{** loaded for the language `#1'. Using the pattern for}%
\typeout{** the default language instead.}%
\else
\language=\csname l@#1\endcsname
\fi
#2}}
\providecommand{\BIBdecl}{\relax}
\BIBdecl

\bibitem{Knill_Laflamme}
E.~Knill and R.~Laflamme, ``Theory of quantum error-correcting codes,''
  \emph{Physical Review A}, vol.~55, no.~2, p. 900, 1997.

\bibitem{Rains}
E.~M. Rains, ``Nonbinary quantum codes,'' \emph{IEEE Transactions on
  Information Theory}, vol.~45, no.~6, pp. 1827--1832, 1999.

\bibitem{Cerf_Cleve}
N.~J. Cerf and R.~Cleve, ``Information-theoretic interpretation of quantum
  error-correcting codes,'' \emph{Physical Review A}, vol.~56, no.~3, p. 1721,
  1997.

\bibitem{Grassl_Huber_Winter}
M.~Grassl, F.~Huber, and A.~Winter, ``Entropic proofs of singleton bounds for
  quantum error-correcting codes,'' \emph{IEEE Transactions on Information
  Theory}, vol.~68, no.~6, pp. 3942--3950, 2022.

\bibitem{Grassl_Beth_Roetteler}
M.~Grassl, T.~Beth, and M.~Roetteler, ``On optimal quantum codes,''
  \emph{International Journal of Quantum Information}, vol.~2, no.~01, pp.
  55--64, 2004.

\bibitem{Huber_Grassl}
F.~Huber and M.~Grassl, ``Quantum codes of maximal distance and highly
  entangled subspaces,'' \emph{Quantum}, vol.~4, p. 284, 2020.

\bibitem{Alsina_Razavi}
D.~Alsina and M.~Razavi, ``Absolutely maximally entangled states,
  quantum-maximum-distance-separable codes, and quantum repeaters,''
  \emph{Physical Review A}, vol. 103, no.~2, p. 022402, 2021.

\bibitem{Facchi_Florio_Parisi_Pascazio}
P.~Facchi, G.~Florio, G.~Parisi, and S.~Pascazio, ``Maximally multipartite
  entangled states,'' \emph{Physical Review A—Atomic, Molecular, and Optical
  Physics}, vol.~77, no.~6, p. 060304, 2008.

\bibitem{Helwig_Cui}
W.~Helwig, W.~Cui, J.~I. Latorre, A.~Riera, and H.-K. Lo, ``Absolute maximal
  entanglement and quantum secret sharing,'' \emph{Physical Review A—Atomic,
  Molecular, and Optical Physics}, vol.~86, no.~5, p. 052335, 2012.

\bibitem{Goyeneche_Alsina_Latorre_Riera_Zyczkowski}
D.~Goyeneche, D.~Alsina, J.~I. Latorre, A.~Riera, and K.~{\.Z}yczkowski,
  ``Absolutely maximally entangled states, combinatorial designs, and
  multiunitary matrices,'' \emph{Physical Review A}, vol.~92, no.~3, p. 032316,
  2015.

\bibitem{Scott}
A.~J. Scott, ``Multipartite entanglement, quantum-error-correcting codes, and
  entangling power of quantum evolutions,'' \emph{Physical Review A—Atomic,
  Molecular, and Optical Physics}, vol.~69, no.~5, p. 052330, 2004.

\bibitem{Raissi_Gogolin_Riera_Acin}
Z.~Raissi, C.~Gogolin, A.~Riera, and A.~Ac{\'\i}n, ``Optimal quantum error
  correcting codes from absolutely maximally entangled states,'' \emph{Journal
  of Physics A: Mathematical and Theoretical}, vol.~51, no.~7, p. 075301, 2018.

\bibitem{Schumacher_Nielsen}
B.~Schumacher and M.~A. Nielsen, ``Quantum data processing and error
  correction,'' \emph{Physical Review A}, vol.~54, no.~4, p. 2629, 1996.

\bibitem{Nielsen_Chuang}
M.~A. Nielsen and I.~L. Chuang, \emph{Quantum computation and quantum
  information}.\hskip 1em plus 0.5em minus 0.4em\relax Cambridge university
  press, 2010.

\bibitem{Pippenger}
N.~Pippenger, ``The inequalities of quantum information theory,'' \emph{IEEE
  Transactions on Information Theory}, vol.~49, no.~4, pp. 773--789, 2003.

\bibitem{Preskill}
J.~Preskill, ``Lecture notes for physics 229: Quantum information and
  computation,'' 1998.

\bibitem{Aharonov_Ben}
D.~Aharonov and M.~Ben-Or, ``Fault-tolerant quantum computation with constant
  error,'' in \emph{Proceedings of the twenty-ninth annual ACM symposium on
  Theory of computing}, 1997, pp. 176--188.

\bibitem{Cleve_Gottesman_Lo}
R.~Cleve, D.~Gottesman, and H.-K. Lo, ``How to share a quantum secret,''
  \emph{Physical review letters}, vol.~83, no.~3, p. 648, 1999.

\end{thebibliography}

\section*{Appendix}
We complement Theorem \ref{thm} with a concrete quantum MDS code construction and verify that it indeed achieves the entropy value in (\ref{entropy}). We use the quantum analogue of Reed Solomon code, which has been applied to fault tolerant quantum computing  \cite{Aharonov_Ben} and quantum secret sharing \cite{Cleve_Gottesman_Lo} in the literature, and present it with Vandermonde matrices (instead of polynomials as in \cite{Aharonov_Ben, Cleve_Gottesman_Lo}).

\vspace{0.1in}
The quantum message $Q_0$ has $k$ qudits where each qudit is $q$-dimensional. Set $q$ as any prime power such that $q \geq n = k + 2(d-1)$. $Q_0$ is maximally mixed and has a purification $R Q_0 = \sum_{a_1, \cdots, a_k \in \mathbb{F}_q} \frac{1}{\sqrt{q^k}} \ket{a_1, \cdots, a_k} \ket{a_1, \cdots, a_k} $. To perform encoding, we append $2(d-1)$ ancilla qudits $Q_{anc} = \sum_{b_1, \cdots, b_{d-1} \in \mathbb{F}_q} \frac{1}{\sqrt{q^{d-1}}} \ket{b_1, \cdots, b_{d-1}, 0, \cdots, 0}$ and proceed as follows.
\begin{eqnarray}
R Q_0 Q_{anc} &=& \sum_{a_1, \cdots, a_k \in \mathbb{F}_q} \frac{1}{\sqrt{q^k}} \ket{a_1, \cdots, a_k} \ket{a_1, \cdots, a_k} \notag\\
&&~~\otimes \sum_{b_1, \cdots, b_{d-1} \in \mathbb{F}_q} \frac{1}{\sqrt{q^{d-1}}} \ket{b_1, \cdots, b_{d-1}, 0, \cdots, 0} \\
&\rightsquigarrow& \sum_{a_1,\cdots, a_k} \frac{1}{\sqrt{q^k}} \ket{a_1, \cdots, a_k} \sum_{b_1,\cdots, b_{d-1}} \frac{1}{\sqrt{q^{d-1}}} \ket{ (a_1, \cdots, a_k, b_1, \cdots, b_{d-1}) ({\bf A}; {\bf B})} \label{enc} \\
&=& R Q_1 \cdots, Q_n
\end{eqnarray} 
where 
$({\bf A}; {\bf B})$ is set as the Vandermonde matrix and represents the vertical concatenations of matrices ${\bf A}$ and ${\bf B}$, i.e., 
\begin{eqnarray}
&& ({\bf A}; {\bf B}) = \left[
\begin{array}{cccc}
\alpha_1^{k+d-2} & \alpha_2^{k+d-2} & \cdots & \alpha_n^{k+d-2} \\
\vdots & \vdots & \ddots & \vdots \\
\alpha_1 & \alpha_2 & \cdots & \alpha_n\\
1 & 1 & \cdots & 1
\end{array}
\right] \in \mathbb{F}_q^{(k+d-1) \times n}, \notag \\
&& \alpha_1, \cdots, \alpha_n ~\mbox{are distinct elements in}~\mathbb{F}_q \label{Van}
\end{eqnarray} 
and ${\bf A} \in \mathbb{F}_q^{k \times n}$ is the first $k$ rows and ${\bf B} \in \mathbb{F}_q^{(d-1) \times n}$ is the last $d-1$ rows. In (\ref{enc}), `$\rightsquigarrow$' denotes a unitary transformation, which holds because the Vandermonde matrix $({\bf A}; {\bf B})$ has full rank.

\vspace{0.1in}
Next consider decoding. We show that for any $\mathcal{I} \subset [n], |\mathcal{I}| = n - (d - 1) = k+d-1$, we may recover the source qudits from $Q_\mathcal{I}$. For a matrix ${\bf A}$, ${\bf A}_{\mathcal{I}}$ represents the sub-matrix of ${\bf A}$ with columns in the index set $\mathcal{I}$. To simplify the notation, denote ${\bf a} = (a_1, \cdots, a_k)$ and ${\bf b} = (b_1. \cdots, b_{d-1})$.
\begin{eqnarray}
R Q_{\mathcal{I}} Q_{\mathcal{I}^c} &=& \sum_{{\bf a}} \frac{1}{\sqrt{q^k}} \ket{{\bf a}} \sum_{{\bf b}} \frac{1}{\sqrt{q^{d-1}}} \ket{ ({\bf a}, {\bf b}) ({\bf A}_{\mathcal{I}}; {\bf B}_{\mathcal{I}})} \ket{ ({\bf a}, {\bf b}) ({\bf A}_{\mathcal{I}^c}; {\bf B}_{\mathcal{I}^c}) } \label{enc1} \\
&\rightsquigarrow& \sum_{{\bf a}} \frac{1}{\sqrt{q^k}} \ket{{\bf a}} \sum_{{\bf b}} \frac{1}{\sqrt{q^{d-1}}} \ket{ ({\bf a}, {\bf b})  } \ket{ ({\bf a}, {\bf b}) ({\bf A}_{\mathcal{I}^c}; {\bf B}_{\mathcal{I}^c})} \label{d1} \\
&\rightsquigarrow& \sum_{{\bf a}} \frac{1}{\sqrt{q^k}} \ket{{\bf a}} \sum_{{\bf b}}  \frac{1}{\sqrt{q^{d-1}}} \ket{{\bf a}}  \ket{ ({\bf a}, {\bf b}) ({\bf A}_{\mathcal{I}^c}; {\bf B}_{\mathcal{I}^c}) } \ket{ ({\bf a}, {\bf b}) ({\bf A}_{\mathcal{I}^c}; {\bf B}_{\mathcal{I}^c})} \label{d2} \\
&=& \sum_{{\bf a}} \frac{1}{\sqrt{q^k}} \ket{{\bf a}, {\bf a}} \sum_{{\bf b}}  \frac{1}{\sqrt{q^{d-1}}} \ket{ ({\bf a}, {\bf b}) ({\bf A}_{\mathcal{I}^c}; {\bf B}_{\mathcal{I}^c}) } \ket{ ({\bf a}, {\bf b}) ({\bf A}_{\mathcal{I}^c}; {\bf B}_{\mathcal{I}^c})} \label{d3} \\
&=& \sum_{{\bf a}} \frac{1}{\sqrt{q^k}} \ket{{\bf a}, {\bf a}} \sum_{{\bf b}'} \frac{1}{\sqrt{q^{d-1}}} \ket{  {\bf b}' , {\bf b}'} \label{d4} \\
&=& R \hat{Q}_0 \otimes \cdots \label{d5}
\end{eqnarray} 
where (\ref{d1}) is unitary because $({\bf A}_\mathcal{I}; {\bf B}_\mathcal{I}) \in \mathbb{F}_q^{(k+d-1) \times (k+d-1)}$ is a full-size square sub-matrix of a Vandermonde matrix and has full rank. (\ref{d2}) is unitary because ${\bf B}_{\mathcal{I}^c} \in \mathbb{F}_q^{(d-1) \times (d-1)}$ is itself a square Vandermonde matrix (refer to (\ref{Van})) thus has full rank and $({\bf a}, {\bf b})$ is invertible to $({\bf a}, ({\bf a}, {\bf b})({\bf A}_{\mathcal{I}^c}; {\bf B}_{\mathcal{I}^c}))$. In (\ref{d4}), we define ${\bf b}' = ({\bf a}, {\bf b})({\bf A}_{\mathcal{I}^c}; {\bf B}_{\mathcal{I}^c})$ and we can change the sum over all possible values of ${\bf b}$ to the sum over all possible values of ${\bf b}'$ because for any fixed ${\bf a}$, when ${\bf b}$ takes all values from $\mathbb{F}_q^{d-1}$, ${\bf b}'$ is invertible to ${\bf b}$ and also takes all values from $\mathbb{F}_q^{d-1}$ (${\bf a} {\bf A}_{\mathcal{I}^c}$ may be viewed as a constant shift term to ${\bf b} {\bf B}_{\mathcal{I}^c}$, which takes all possible values). Therefore in the end (\ref{d5}), we have perfectly recovered all source qudits (along with the entanglement with the reference system) as $R \hat{Q}_0$ is now unentangled with the rest of the system.

\vspace{0.1in}
Finally, we compute the entropy values of all sub-systems of the pure coded state (\ref{enc1}) and verify that (\ref{q1}), (\ref{q2}), (\ref{r1}), (\ref{r2}) hold. To this end, we use Lemma \ref{pureh}, presented in the subsection below and the problem reduces to that of computing the dimension of the intersection of any sub-system and its complement.
\begin{eqnarray}
(\ref{q1}): && |\mathcal{I}| \leq k+d-1 \\
&& \mbox{In}~R Q_{\mathcal{I}^c}, {\bf a}, ({\bf a}, {\bf b}) ({\bf A}_{\mathcal{I}^c}; {\bf B}_{\mathcal{I}^c})~\mbox{may recover ${\bf a}, {\bf b}$}.\\
&& \mbox{So the intersection has dimension $|\mathcal{I}|$ and}~H(Q_\mathcal{I}) = |\mathcal{I}|.\\
(\ref{q2}): && |\mathcal{I}| > k+d-1 \\
&& \mbox{In}~Q_{\mathcal{I}}, ({\bf a}, {\bf b}) ({\bf A}_{\mathcal{I}}; {\bf B}_{\mathcal{I}})~\mbox{may recover ${\bf a}, {\bf b}$}.\\
&& \mbox{So the intersection has dimension $k + |\mathcal{I}^c| = k + n - |\mathcal{I}|$ and}\notag\\
&& H(Q_\mathcal{I}) = k + n - |\mathcal{I}| = 2(k+ d-1) - |\mathcal{I}|.
\end{eqnarray}
(\ref{r1}) and (\ref{r2}) follow as the joint state is pure (the above procedure will work equally well). The proof is thus complete.

\subsection*{Entropy of a Pure Uniform Superposition State}
Consider a $1 \times m$ row vector ${\bf x} = (x_1, x_2, \cdots, x_m)$ where each $x_i$ is from finite field $\mathbb{F}_q$ and an $m \times l$ matrix over $\mathbb{F}_q$, ${\bf H} \in \mathbb{F}_q^{m \times l}$ where $\mbox{rank}({\bf H}) = m$. Further, ${\bf H}$ has a column bipartition as ${\bf H} = ( {\bf H}_1, {\bf H}_2)$ where ${\bf H}_1 \in \mathbb{F}_q^{m \times l_1}$, ${\bf H}_2 \in \mathbb{F}_q^{m \times l_2}$, and $l_1 + l_2 = l$. For a matrix ${\bf H}$ over $\mathbb{F}_q$, let $\langle {\bf H} \rangle$ represent the vector space spanned by the columns of ${\bf H}$ over $\mathbb{F}_q$.
The entropy of the bipartition of a pure quantum state generated by uniform superposition over all possible values of ${\bf x}$ 
is characterized in the following lemma.

\begin{lemma}\label{pureh}
Consider the following pure uniform superposition state of $l$ qudits, where each qudit is $q$-dimensional, $\ket{\psi} = \frac{1}{\sqrt{q^m}} \sum_{ {\bf x} \in \mathbb{F}_q^m} \ket{{\bf x} {\bf H}^{m \times l}} = A B$, where $A$ denotes the sub-system of $\ket{\psi}$ that consists of the first $l_1$ qudits and $B$ denotes the sub-system of $\ket{\psi}$ that consists of the last $l_2$ qudits. Then the entropy of $H(A)_{\ket{\psi}}$ measured in $q$-ary units is 
\begin{eqnarray}
H(A) = H(B) = \mbox{dimension}(\langle {\bf H}_1 \rangle \cap \langle {\bf H}_2 \rangle).
\end{eqnarray}
\end{lemma}

\proof Suppose $\mbox{dimension}(\langle {\bf H}_1 \rangle) = \delta_1$, $\mbox{dimension}(\langle {\bf H}_2 \rangle) = \delta_2$, and $\mbox{dimension}(\langle {\bf H}_1 \rangle \cap \langle {\bf H}_2 \rangle) = \delta_{12} = \delta_1+\delta_2 - m$. We may find matrices ${\bf B}_{12} \in \mathbb{F}_q^{m \times \delta_{12}}$, ${\bf B}_{1} \in \mathbb{F}_q^{m \times (\delta_1 - \delta_{12})}$, ${\bf B}_2 \in \mathbb{F}_q^{m \times (\delta_2 - \delta_{12})}$ 
such that
\begin{enumerate}
\item ${\bf B}_{12}$ is a basis of $\langle {\bf H}_1 \rangle \cap \langle {\bf H}_2 \rangle$
\item $({\bf B}_1, {\bf B}_{12})$ is a basis of $\langle {\bf H}_1 \rangle$
\item $({\bf B}_2, {\bf B}_{12})$ is a basis of $\langle {\bf H}_2 \rangle$
\item 
$({\bf B}_1, {\bf B}_2, {\bf B}_{12})$ is a basis of $\langle {\bf I}_m \rangle$ where ${\bf I}_m$ denotes the $m \times m$ identity matrix
\end{enumerate}
and then we can perform a change of basis operation (invertible matrix multiplication), i.e., there exist full rank matrices ${\bf T}_1 \in \mathbb{F}_q^{l_1 \times l_1}, {\bf T}_2 \in \mathbb{F}_q^{l_2 \times l_2}$ such that 
\begin{eqnarray}
&& {\bf H}_1 {\bf T}_1 = \left({\bf 0}_{m \times (l_1 - \delta_1)}, {\bf B}_1, {\bf B}_{12} \right) \label{t1}\\
&& {\bf H}_2  {\bf T}_2 = \left({\bf 0}_{m \times (l_2 - \delta_2)}, {\bf B}_2, {\bf B}_{12} \right) \label{t2}
\end{eqnarray}
where ${\bf 0}_{a \times b}$ denotes the $a \times b$ matrix with all elements being zero. Define
\begin{eqnarray}
&& \vec{\alpha}_{12} = \big(\alpha_{12}(1), \cdots, \alpha_{12} ({\delta_{12}}) \big) \triangleq {\bf x} {\bf B}_{12} \label{a1}\\
&& \vec{\alpha}_1 = \big(\alpha_1(1), \cdots, \alpha_1({\delta_1 - \delta_{12}}) \big) \triangleq {\bf x} {\bf B}_{1} \label{a2}\\
&& \vec{\alpha}_2 = \big(\alpha_2(1), \cdots, \alpha_2({\delta_2 - \delta_{12}}) \big) \triangleq {\bf x} {\bf B}_{2} \label{a3} 
\end{eqnarray}
and then
\begin{eqnarray}
AB &=&  \frac{1}{\sqrt{q^m}} \sum_{ {\bf x} \in \mathbb{F}_q^m} \ket{{\bf x} {\bf H}^{m \times l}} \\
&=& \frac{1}{\sqrt{q^m}} \sum_{ {\bf x} \in \mathbb{F}_q^m} \ket{{\bf x} {\bf H}_1^{m \times l_1}}  \ket{{\bf x} {\bf H}_2^{m \times l_2}} \\
&\rightsquigarrow& \frac{1}{\sqrt{q^m}} \sum_{ {\bf x} \in \mathbb{F}_q^m} \ket{{\bf x} \left({\bf 0}_{m \times (l_1 - \delta_1)}, {\bf B}_1, {\bf B}_{12}\right)}  \ket{{\bf x} \left({\bf 0}_{m \times (l_2 - \delta_2)}, {\bf B}_2, {\bf B}_{12} \right) } \label{l1}\\
&=& \frac{1}{\sqrt{q^m}} \sum_{ {\bf x} \in \mathbb{F}_q^m} \ket{{\bf 0}_{1\times(l_1-\delta_1)} , \vec{\alpha}_1, \vec{\alpha}_{12}} \ket{{\bf 0}_{1\times(l_2-\delta_2)} , \vec{\alpha}_2, \vec{\alpha}_{12}} \label{l2}\\
&=& \frac{1}{\sqrt{q^m}} \sum_{ (
\vec{\alpha}_1, \vec{\alpha}_2, \vec{\alpha}_{12}) \in \mathbb{F}_q^m} \ket{{\bf 0}_{1\times(l_1-\delta_1)} , \vec{\alpha}_1, \vec{\alpha}_{12}} \ket{{\bf 0}_{1\times(l_2-\delta_2)} , \vec{\alpha}_2, \vec{\alpha}_{12}} \label{l3} \\
&\rightsquigarrow& \frac{1}{\sqrt{q^{m}}} 
\underbrace{ \ket{{\bf 0}_{1\times(l_1-\delta_1)} }}_{A_0}  \otimes \ket{{\bf 0}_{1\times(l_2-\delta_2)} }   \otimes \underbrace{ \sum_{ \vec{\alpha}_1} \ket{\vec{\alpha}_1}}_{A_1} \otimes \sum_{ \vec{\alpha}_2} \ket{\vec{\alpha}_2} \otimes \sum_{ \vec{\alpha}_{12}} \underbrace{\ket{\vec{\alpha}_{12}}}_{A_{12}} \underbrace{\ket{\vec{\alpha}_{12}}}_{B_{12}} \label{l4}
\end{eqnarray}
where (\ref{l1}) is a unitary transformation because ${\bf T}_1, {\bf T}_2$ have full rank (refer to (\ref{t1}), (\ref{t2})). To obtain (\ref{l2}), we plug in the definition of $\vec{\alpha}_1, \vec{\alpha}_2, \vec{\alpha}_{12}$ (refer to (\ref{a1}), (\ref{a2}), (\ref{a3})). In (\ref{l3}), we may replace the sum over ${\bf x} \in \mathbb{F}_q^{m}$ to the sum over $(\vec{\alpha}_1, \vec{\alpha}_2, \vec{\alpha}_{12}) \in \mathbb{F}_q^{\delta_1+\delta_2-\delta_{12}}$ (note that $m = \delta_1 + \delta_2 - \delta_{12}$) because ${\bf x}$ is invertible to $(\vec{\alpha}_1, \vec{\alpha}_2, \vec{\alpha}_{12})$ (note that $({\bf B}_1, {\bf B}_2, {\bf B}_{12})$ has full rank). In (\ref{l4}), we reorder the qudits and separate out the unentangled parts, then $A$ is divided into three parts, $A_0, A_1, A_{12}$. We are now ready to compute the entropy of $A$. Note that unitary transformations do not change entropy.
\begin{eqnarray}
H(A) &=& H(A_0, A_1, A_{12}) \label{l5}\\
&=& H(A_0) + H(A_1) + H(A_{12}) \label{l6} \\
&=& 0 + 0 + \delta_{12} \label{l8}\\
&=& \mbox{dimension}(\langle {\bf H}_1 \rangle \cap \langle {\bf H}_2 \rangle) \label{l7} \\
&=& H(B)
\end{eqnarray}
where 
(\ref{l6}) is due to the fact that $A_0, A_1, A_{12}$ are in a product state (refer to (\ref{l4})). To obtain (\ref{l8}), we use the fact that $A_0, A_1$ are pure and $A_{12}$ is maximally entangled with $B_{12}$ (refer to (\ref{l4})).

\hfill\QED

\end{document}